\documentclass[prd,aps,floats,preprintnumbers]{article}
\usepackage{graphicx, epsfig}
\usepackage{latexsym}
\usepackage{amsthm}
\usepackage{amsmath}   
\usepackage{amssymb}
\newcommand{\beeqnar}{\begin{eqnarray}}
\newcommand{\en}{\end{eqnarray}}
\renewcommand{\L}{\mathcal{L}}
\newcommand{\nn}{\nonumber}
\newcommand{\ep}{\epsilon}
\newcommand{\M}{\mathcal{M}}
\renewcommand{\O}{\mathcal{O}}
\newcommand{\dd}[2]{\frac{\partial #1}{\partial #2}}
\renewcommand{\d}{\partial}

\newcommand{\be}{\begin{equation}}
\newcommand{\ee}{\end{equation}}
\newcommand{\bea}{\begin{eqnarray}}
\newcommand{\eea}{\end{eqnarray}}

\newcommand{\half}{\frac{1}{2}}

\begin{document}

%\preprint{MAD-TH-03-2, MAD-PH-03-XX}

\setlength{\unitlength}{1mm}
 
\begin{titlepage}

%MAD-TH-03-2, MAD-PH-03-XX

\title{Searching for the Kaluza-Klein Graviton in Bulk RS Models}
 
\author{Liam Fitzpatrick$^1$, Jared Kaplan$^1$, Lisa Randall$^1$,
  Lian-Tao Wang$^{1,2}$ \\ 
{\it $^1$Department of Physics, Harvard University, Cambridge, MA
  02138  } \\ 
{\it $^2$Department of Physics, Princeton University, Princeton, NJ. 08544}}
\date{\today}
\maketitle

\begin{abstract}
The best-studied version of the RS1 model has all the Standard Model particles 
confined to the TeV brane. However,  recent variants have 
the Standard Model fermions and gauge bosons located in the bulk 
five-dimensional spacetime. We study the potential reach of the LHC in 
searching for the lightest KK partner of the graviton in the most
promising such models in which the right-handed top 
 is localized very near the TeV brane and the light fermions
are localized near the Planck brane. We consider both 
detection and the establishment of the spin-2 nature of the resonance should 
it be found.
\end{abstract}

%\pacs{PACS Numbers: }
 
%\newpage

\end{titlepage}

\section{Introduction}

RS models with Standard Model particles in the bulk are a viable
possibility for two reasons. The first is that the solution to the
hierarchy problem requires only that the Higgs particle is on the TeV
brane. Standard Model particles can be either in the bulk or on the
brane. The second important feature of RS1 is that the
higher-dimensional space, the fourth spatial dimension, is quite
small, only of order 35 times the AdS length. Because of this feature,
gauge bosons can be in the bulk without the coupling being too strong,
since the forces are not very highly diluted. 
Such models, with the Standard Model in the bulk and
only the Higgs
sector on the TeV brane, have phenomenological advantages that include
possibilities for avoiding precision constraints from light quark
interactions, allowing high-scale 
unification of gauge couplings, and a natural hierarchy of masses\cite{rs}.

However, studies of the phenomenological consequences of the
Kaluza-Klein mode of 
the graviton in RS theories
have focused primarily on  the scenario where all Standard Model matter resides
on the TeV brane (e.g. \cite{DHRbrane,DHRphotons}), or where the Standard Model
gauge fields and fermions are in the bulk but the top and lighter fermions 
are all localized at the same place in the bulk \cite{DHR,pomarolRSbulk}.
In more realistic models, top quarks are localized near the TeV brane and
right-handed isospin \cite{agashe} is gauged.  In these models, 
precision electroweak constraints are weaker with the consequence that
a new decay signature - KK graviton decay to tops - becomes 
significant.  Previously, electoweak constraints put the graviton mass
above 23 TeV for tops localized very near the TeV brane\cite{DHR}.
The weakened constraints from the specific model \cite{agashe} 
are almost (but not quite) weak enough to allow an observable KK graviton,
and we expect that a modest amount of model-building could
lower them further.

If indeed the Standard Model fields  are in the bulk, the first set of
resonances to be discovered will most likely be  
KK-gluons \cite{kkgluon}, and possibly other spin-1 KK excitations of
the SM gauge 
boson.  However, although the spin-1 resonance would be quite an
exciting discovery, it will not be sufficient to determine the
underlying nature of the model. It will have the properties of a
resonance from a strongly  interacting theory that is coupled
primarily to the right-handed top quark, and it might not be readily
distinguished on its own from a purely four-dimensional
model. Discovery of the spin-2 resonance, though not conclusive
either, will demonstrate that a Randall-Sundrum type of setup is a
more likely description of new physics.   

In this paper, we set out to study the phenomenology of the RS KK
graviton when the light quarks are localized near the Planck brane but the
top quarks are localized very near the TeV brane, as would be expected
in any bulk model which with a sufficiently large top quark mass. 
We will discuss the collider reach of the KK resonance, as well
as ways of determining its properties. We will not make any assumptions about
the minimum value of the KK mass (as are implied by electroweak
constraints) but will simply leave the mass as a free parameter to see the
sensitivity of direct KK graviton searches. 
 In calculating the collider reach, we assume 100 fb$^-1$ luminosity and perfect 
top tagging. For realistic case with some top tag efficiency and potentially 
with larger luminosity, the estimate of reach could be obtained from our study 
easily by scaling.

We find that the primary production mode is through KK gluon
annihilation, which can lead to measurable KK graviton resonances up
to about 1.4 TeV, about one-quarter the mass reach of the model with SM
particles confined to the TeV brane.  However, the angular
distribution of the decay products when the KK graviton is produced
through the annihilation of two spin-1 gluons is quite distinctive,
and should allow for angular determination with fewer particles than
would be necessary in the model with SM particles bound to the brane.
Furthermore, even at large values of the  
AdS curvature scale, approaching the Planck scale, we find that the KK graviton
has a very narrow width.  This is not true in models with fermions on
the brane, 
since in that case the KK graviton can decay to a large number of light
fermion degrees of freedom.  This distinctive feature should be an
advantage that partially compensates for the lower production rate of
the KK gravitons in any given mass range. 

\section{Production and decay}

We are interested in the production and decay of KK gravitons in an RS model where all 
Standard Model fields except for the Higgs boson propagate in the bulk, and the light fermions are localized near the Planck brane whereas the right-handed top quark is localized near the TeV brane. The KK graviton
is also localized near the TeV brane, and in the dual CFT this means that it is a composite 
state with the large N scaling properties of a glueball.  Thus it couples most strongly
to other composite states -- the Higgs boson, the top quark, other KK modes, and to a 
lesser extent, gauge boson zero modes, whose interactions are suppressed from the 5d
point of view by the volume of the fifth dimension.
  The KK graviton couples very weakly to light
quarks and leptons, because they are localized near the Planck brane (or in the CFT, they 
are elementary fields) in order to explain their small Yukawa couplings and masses
and also to shield the theory from large operators violating precision
electroweak constraints.

Thus we expect that hadron colliders will produce KK gravitons through
gluon annihilation, 
and that these gravitons will decay to Higgs bosons, Ws, Zs, top
quarks, and other KK states, the lightest of which are always lighter
than the gravitons.   Although production through
WW fusion might be possible, we find that it is numerically smaller
than production through gluons.

\subsection{Setup}

In this section, we derive the relevant interaction terms.  The 
coupling constants depend on the overlap of the particle wavefunctions
in the extra dimension, and for further details of these calculations
we refer the reader to \cite{DHR}.  We take $k=1/L$ to be the AdS curvature
scale and $r_c$ the proper size of the extra dimension.  The 
effective cut-off on the theory is then $\mu_{TeV} = k e^{-\pi k r_c}$,
and all interactions with the KK graviton are suppressed by
this scale.  $k$ is taken near the Planck scale $M_4 \equiv
\frac{1}{\sqrt{8 \pi G_N}}$, and we leave their ratio $M_4/k = M_4 L$
a free parameter unless otherwise stated.  $M_4 L$ is
proportional to the $N$ of the dual CFT.
  We define $\nu = m/k$, where $m$ is the bulk mass for fermion fields; 
this parameter determines where the lightest mode of the fermion is localized in the bulk, or equivalently,
its admixture of composite CFT states.  We will refer to the localization
parameter $\nu$ for the top-right quark as $\nu_{t,R}$, and in order to get a 
heavy top quark, $\nu_{t,R}$ will be greater than $\nu$ for the lighter fermions.
Although we are motivated by the Agashe et. al. paper, 
we are not confined by their parameter choice and will  allow $nu_{t,R}$ 
to vary over a wide range.  
All couplings to the KK graviton 
can be written in the form \cite{DHR}
\begin{equation}
C_{XXG} \int d^4 x h_{\mu \nu} T^{\mu \nu}_{XX} 
\end{equation}
where $XX$ indicates either a pair of fermions or gauge fields, 
$h_{\mu \nu}$ is the 
field for the KK-graviton, and $T^{\mu \nu}$ is the effective 4-d energy 
momentum tensor.  The relevant couplings for KK graviton production and
decay are the graviton coupling to two zero-mode gluons, to 
a top-anti-top pair, to two scalars, and to a top and a KK anti-top.  
In each case, the stress-energy tensor $T_{\mu\nu}$ takes the form
\begin{equation}
T_{\mu\nu} = 2 \dd{\L}{g^{\mu\nu}} - g_{\mu\nu} \L
\end{equation}
and the $g_{\mu\nu} \L$ piece can be ignored since the KK graviton
polarizations are traceless. 

The mass spectrum of the KK modes also enters the form of the couplings.
The KK masses take the form $x_n^X \mu_{TeV}$, where $x_n^X$ is a root
of the boundary condition for the specific particle.  For gravitons, the boundary
condition is $J_1(x)=0$, and the first root is $x_1^G = 3.83$.  For
gauge bosons when $k \pi r_c = 35$, the first root is $x_1^A = 2.45$.  
For fermions,
the boundary condition is
\begin{equation}
\frac{J_{-(\nu + \half)} (x_n^L \epsilon)}{Y_{-(\nu+\half)}(x_n^L \epsilon)}
= \frac{J_{-(\nu + \half)} (x_n^L)}{Y_{-(\nu+\half)}(x_n^L)}
\end{equation}
For $\nu > -1/2$, this condition is approximately $tan( \nu \pi) = 
tan ( x_n^L + \nu \pi/2)$, 
which implies 
\begin{equation}
x_n^L \approx \pi ( n + \nu/2) \qquad (\nu>-1/2)
\label{kkfermionspectrum}
\end{equation}

\begin{table}
\begin{tabular}{|c|c|l|}
\hline
$XX$ & $T_{XX}^{\mu\nu}$ & $c_{XXG}$ \\
\hline
$ss$ (scalars) & $\half \d^\mu\phi \d^\nu \phi $ & 
   $c_{ssG} = \frac{2}{(M_4 L) \mu_{TeV}}$ \\
\hline
$f\bar{f}$ (fermions) & $i \psi^\dagger \bar{\sigma}^\mu D^\nu \psi$
  & $c_{f\bar{f} G} =   
\frac{1}{(M_4 L) \mu_{TeV}} 
\left( \frac{1 + 2 \nu}{1 - e^{-\pi k r_c (1 + 2 \nu)}} \right)
\frac{\int_0^1 dy \ y^{2 + 2 \nu} J_2(3.83 y) }{J_2(3.83)} $\\
\hline
$t\bar{t}_1$ (top+KK-top) & $i \psi^\dagger \bar{\sigma}^\mu D^\nu \psi$ & 
$c_{f\bar{f}G}^{101} = \frac{1}{(M_4 L) \mu_{TeV} } \sqrt{\frac{2 (1+2 \nu)}
{1-\epsilon^{1+2\nu}}} \int_0^1 dy y^{\nu + 5/2} 
\frac{J_{\nu-1/2} (x_1^L y)}{J_{\nu-1/2}(x_1^L)} 
\frac{J_2(3.83 y)}{|J_2(3.83)|}$\\
\hline
$gg$ (gluons) & $F^{\mu \rho} F^\nu_{\ \rho}$ & $ 
c_{ggG} = \frac{1}{(\pi k r_c)(M_4 L) \mu_{TeV}} 
\frac{\int_0^1 dy \ y J_2(3.83 y) }{J_2(3.83)}
\approx \frac{0.47}{(\pi k r_c)(M_4 L) \mu_{TeV}} $ \\
\hline
\end{tabular}
\label{vertices}\caption{All couplings to the graviton are of the form
$c_{XXG} h_{\mu\nu}T^{\mu\nu}_{XX}$.  Terms in $T_{\mu\nu}$ 
proportional to $\eta_{\mu\nu}$ have been dropped, since $h_{\mu\nu}$
is traceless.}
\end{table}

We collect the interactions together in Table \ref{vertices}.  The
$\eta_{\mu\nu} \L$ term is dropped in the expression for $T_{\mu\nu}$.
Some of the qualitative features of the couplings are readily understandable.
In all the couplings, the suppression by $(M_4 L) \propto N$ arises because 
the
KK graviton has the $N$ scaling of a glueball state.  
The factor of $1/\mu_{TeV}$ is the 
local $UV$ scale, and it serves as a cutoff.
  The suppression by $\pi k r_c$ in $c_{ggG}$ 
follows because the gauge field has a flat wave function, indicating that 
its couplings to the brane-localized KK graviton modes are suppressed
by the volume of the bulk. 
The fermion coupling $c_{f\bar{f}G}$ has a strong dependence on whether
the fermion bulk wavefunction is localized near the TeV brane or the 
Planck brane.  This dependence is contained in the factor in parentheses 
(the bessel function integral only varies by about a 
factor of 2 as $\nu$ varies from $-1$ to $1$).  
Note that for generic $\nu > -1/2$, relevant for heavy fermions like the top,
which are located near the TeV brane, the factor in parentheses
is of order one. For $\nu$ very close to $-1/2$ it is 
approximately $1/(\pi k r_c)$, again the volume suppression of a flat wave 
function. For $\nu < -1/2$, relevant for light fermions near the 
Planck brane,
it is exponentially small.  

The coefficient $c_{ssG}$ for a scalar on the TeV brane is 
relevant to both the Higgs and the longitudinal 
components of the W and Z. This follows from the Goldstone Boson
Equivalence theorem, as we review in appendix
\ref{Goldstone}.

\subsection{Electroweak Constraints}

A specific example is provided by the model of Agashe et al \cite{agashe},
where an additional gauged $SU(2)$ isospin in the bulk suppresses contributions
to the Peskin-Takeuchi $T$ parameter.  In this model, the constraints from
the $S$- and $T$-parameters are of roughly equal importance,  the contribution
to $S$ is
\begin{equation}
S = 2 \pi \left( \frac{v}{\mu_{TeV}} \right)^2 = 
  0.20 \left( \frac{ 5.15 \ \mathrm{TeV}}{m_{grav}} \right)^2 .
\end{equation}
However, note that the second equality follows from the tree level relation between the
KK-graviton and KK-gauge boson masses in this particular model, so \emph{the constraint
on the mass of the KK graviton is indirect}.
% XXX SENTENCE ADDED PER REF'S REQUEST -- WANTED US TO CLARIFY THIS POINT
The Higgs makes an additional positive contribution to the $S$ parameter.  The
$1\sigma$ error on $S$ is about 0.10 \cite{pdg:ew}.  In this model $m_{grav} \lesssim 5$ TeV 
will result in too large a value for $S$.  % XXX MINOR CHANGE IN WORDING, "RESULT IN" FOR "GIVE"
Negative contributions to $S$ can however partially cancel this contribution, 
permitting lower values for the graviton mass. Also, brane kinetic terms for the
graviton can lower its mass relative to the cut-off scale $\mu_{TeV}$, making
precision electroweak constraints less restrictive for KK graviton phenomenology.  
A substantial brane kinetic term would alter couplings and could lead to interesting
phenomenology, but we will not consider this scenario further.
% XXX ADDED ABOVE PER REF'S REQUEST

In fact, there is some theoretical prejudice for a lower cutoff scale, and therefore
a smaller KK graviton mass, since this sets a minimum level of fine-tuning for the Higgs 
mass.  In particular, if the loop contributions to the Higgs mass are cut off at
the Planck scale, then the Higgs vev is
\begin{equation}
v \sim \frac{(M_4 L) \mu_{TeV}}{\sqrt{2 \lambda}}
\end{equation}
where $\lambda$ is the Higgs quartic coupling.
So, despite the electroweak constraints on the specific model described above,
we will consider $\mu_{TeV}$ as small as 240 GeV, the standard model Higgs vev.  
There are direct constraints that rule out such a small $\mu_{TeV}$, but in order
to be as model independent as possible, we will begin with this theoretically 
motivated minimum value for $\mu_{TeV}$ in our scan of parameter space.

\subsection{Cross Sections}

\begin{figure}
\begin{center}
\includegraphics[angle=270,scale=0.5]{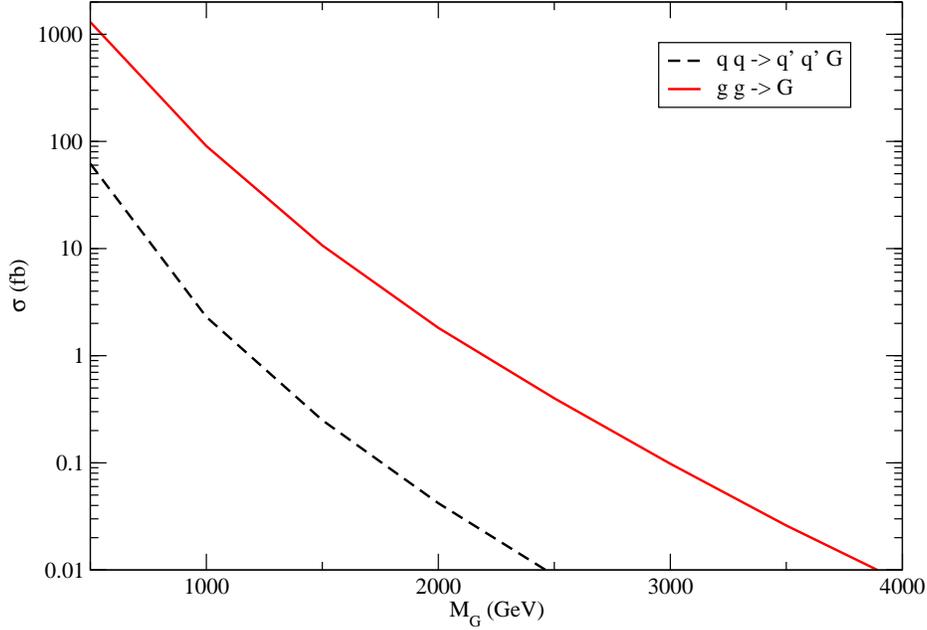}
\end{center}
\caption{Cross section of KK graviton production with $M_4 L=2.5$. Both the 
cross-section from gluon fusion $gg\rightarrow G$ and 
W boson fusion $qq \rightarrow q'q' WW \rightarrow q'q' G$ are shown.
\label{xsec}} 
\end{figure}

The largest contributions to KK graviton production (and decay) come from  $gg \to G \to f \bar{f}$ and $gg \to G \to \phi \phi$,
where the scalar final states are appropriate for either the Higgs boson or for longitudinal
Ws and Zs via the Goldstone boson equivalence theorem.  The KK graviton propagator is
\begin{equation}
D^{\mu \nu , \lambda \sigma} (k) = \frac{G_{\mu \lambda} G_{\nu \sigma} + G_{\mu \sigma} G_{\nu \lambda} - 
\frac{2}{3} G_{\mu \nu} G_{\lambda \sigma} }{2 (k^2 - m^2)}
\end{equation}
with
\begin{equation}
G_{\alpha \beta} = g_{\alpha \beta} - \frac{k_\alpha k_\beta}{m^2}
\end{equation}
Note that $D$ is traceless over $\mu \nu$ and $\lambda \sigma$, so the
KK graviton does not couple 
to the trace of $T_{\mu \nu}$ as expected.  The matrix element for
$gg\rightarrow G \rightarrow t\bar{t}$ can be calculated by contracting
\begin{eqnarray}
\mathcal{M} &=& T_{gluon}^{\mu\nu} D_{\mu\nu, \lambda \sigma}
T_{top}^{\lambda \sigma} 
\end{eqnarray}
We compute the cross section and integrate over phase space numerically.
The resulting cross-section for KK graviton production is shown in 
Fig.~\ref{xsec}.  Note that it gives only the
cross-section for production and does not include an additional branching ratio
for subsequent decays, given by the widths calculated in the next section.
As we can see, the production cross section (assuming 100 $fb^{-1}$)
 peters out at about 4 TeV. We will compare to background in the following 
section to get a better idea of the discovery reach.

   The KK gravitons can also be produced by W boson fusion (WBF), 
though this is only a small fraction of gluon fusion cross-section.
For comparison, the cross-section calculated by Monte Carlo integration 
of the WBF process is also given in Fig.~\ref{xsec}.  

%\begin{table}
%\begin{tabular}{|lll|lll|}
%\hline
%$m_{grav}$ (TeV) & $\sigma_{gg\rightarrow G}$ (fb) &
%$\sigma_{WW\rightarrow G}$ (fb) & $m_{grav}$ (TeV) &
%$\sigma_{gg\rightarrow G}$ (fb) & $\sigma_{WW\rightarrow G}$ (fb) 
%\\
%\hline
%0.5 & 1300 & 62 & 2.5 & 0.40 & 0.0089\\
%1.0 & 90.6 & 2.3 & 3.0 & 0.098 & 0.0022\\
%1.5 & 10.7 & 0.25 & 3.5 & 0.026 & 0.00056\\
%2.0 & 1.82 & 0.042 & 4.0 & 0.0077 & 0.00016\\
%\hline
%\end{tabular}
%\label{crosssection}\caption{Results from numerical calculation
%of cross-section using Monte Carlo integration.  Cross-section
%for W fusion is shown for comparison.}
%\end{table}

\subsection{Decay Rates/Width}

\begin{figure}
\begin{center}
\includegraphics[angle=270,scale=0.5]{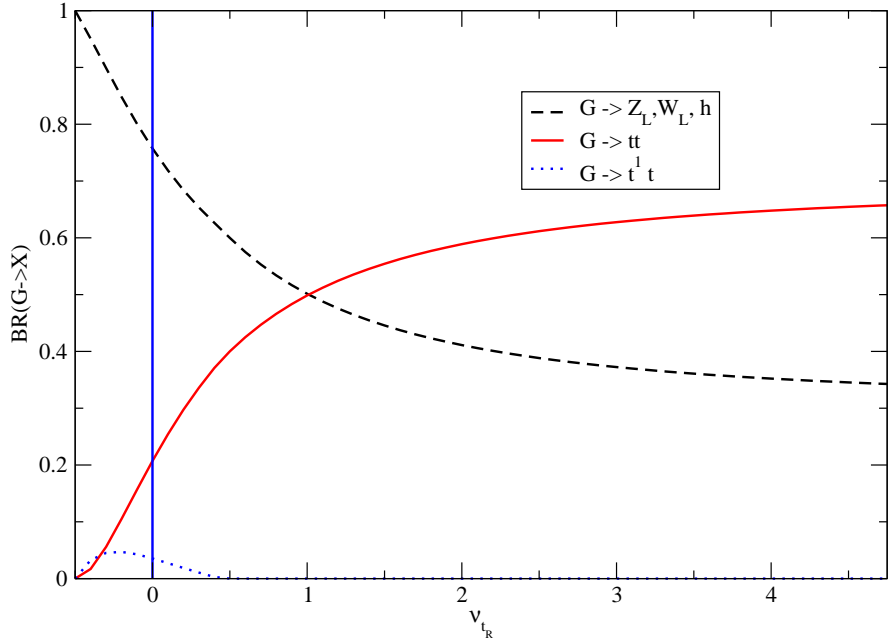}
\end{center}
\caption{Branching Ratios for graviton decay to scalars and
  quarks as a function of the top-right localization parameter $\nu_{t,R}$.
At $-0.5 < \nu_{t,R}< -0.2$, the dominant decay is to the Higgs and 
longitudinal gauge bosons $Z_L,W_L^\pm$. At $\nu_{t,R} > -0.2$, the
dominant decay is to $t\bar{t}$.  The decay to a zero-mode
top and a KK anti-top is kinematically allowed in the range
$-0.5 < \nu_{t,R} < 0.5$. The line at $\nu_{t,R}$ corresponds to the 
specific choice made in ~\cite{agashe}. \label{rate}} 
\end{figure}

The width of KK-graviton is dominated by the top quark (due to the 
large coupling to the right-handed top quark) and the TeV brane scalars 
(the Higgs boson and longitudinal W's and Z's). Other decay modes are 
suppressed by  the volume factor $1/(\pi k r_c)^2 \sim 1/900$.
The total width due to all four real scalar degrees of freedom is
% XXX NEEDS TO BE CORRECTED AND ALSO THE REFEREE WANTS IT TO BE CLEARER THAT
% THIS IS FOR ALL 4 DEGREES OF FREEDOM XXX
\begin{equation}
\Gamma_{\rm Z_L,W_L,h} = \frac{1}{(M_4 L)^2 \mu_{TeV}^2} \frac{m_{grav}^3}{240 \pi}
 = \left( \frac{2.5}{M_4 L} \right)^2 \frac{m_{grav}}{320}
\end{equation}
For the $t_R$ quark contribution to the width, we find
\begin{equation}
\Gamma_{\rm top} = \frac{1}{(M_4 L)^2 \mu_{TeV}^2} \left( \frac{1 + 2
  \nu_{t,R}}{1 - e^{-\pi k r_c (1 + 2 \nu_{t,R} )}} 
\frac{ \int_0^1 dy \ y^{2 + 2 \nu_{t,R}} J_2(3.83 y) }{J_2(3.83)} \right)^2
  \frac{3 m_{grav}^3}{320 \pi} 
\end{equation}
which is about $m_{grav}/320$ for $M_4 L = 2.5$ and $\nu_{t,R}=1$.  The 
branching ratios are plotted in figure \ref{rate}.
% XXX ADDED ABOVE SENTENCE PER REF'S REQUEST

For a range of possible $\nu_{t,R}$, the decay to a KK top and a zero-mode top will
also be allowed.  From equation (\ref{kkfermionspectrum}), 
the mass of the KK top is approximately $(1 + \nu_{t,R}/2) \pi \mu_{TeV}$; the mass
will be less than
the KK graviton mass $3.83 \mu_{TeV}$ for $\nu_{t,R} < 1/2$.  Below 
this value, the decay width is 
\begin{equation}
\Gamma_{t^{1} t} = \frac{1}{(M_4 L)^2 \mu_{TeV}^2} \left( \frac{2(1+2\nu_{t,R})}
{1-\epsilon^{2\nu_{t,R}+1}} \right) \left( \int_0^1 dy y^{\nu_{t,R} + 5/2}
\frac{J_{\nu_{t,R}-1/2}(x_1^L y)}{J_{\nu_{t,R}-1/2}(x_1^L)} \frac{J_2(3.83 y)}{|J_2(3.83)|}
\right)^2  \frac{3 m_{grav}^3 }{320 \pi} \left| \frac{2 p_f}{m_{grav}} \right|
\end{equation}
where $p_f$ is the spatial momentum of either outgoing decay product. 
This kinematic factor vanishes when the zero mode top and KK top 
mass sum to the KK graviton mass, so the decay shuts off at a little below
$\nu_{t,R} = 1/2$.  

Although for small $M_4 L$ the widths become large, this is the most
favorable limit for discovery, because the production rate also grows
in this limit.  In fact, the widths do not even become very large
for small $M_4 L$, and even for $M_4 L = 1$ and $\nu_{t,R} = 2$, the
width is only six percent of the KK graviton mass.  This is distinctly
different from the case when the Standard Model is on the brane and
the graviton can decay to a wide array of light fermions.  With the
Standard Model fermions on the brane, at $M_4 L $ less than about 2, the
KK graviton cross-section flattens out and the tower of KK modes 
blur together, no longer appearing as individual resonances \cite{DHR}.

\section{Discovery Reach}

Before discussing the discovery reach of the KK-graviton, we first 
consider potential backgrounds.  As discussed above, $gg \rightarrow G^{1} \rightarrow t
\bar{t}$ is the dominant mode so   $t \bar{t}$ production from the
Standard Model is the chief background and is the one we consider in our discovery reach estimates below.

Another potential source of background could be  KK gauge
Bosons, which also decay into $t \bar{t}$ final states. The most
important example is the KK-gluon, which  has a much larger production
cross-section and could therefore  be one of the  the major backgrounds through the
KK-gluon $\rightarrow t \bar{t}$ channel. However, we expect the invariant
masses of the two resonances should be quite different. With no brane
kinetic terms for the gauge boson, the masses of the first KK gluon 
resonance and the first KK graviton resonance differ by a factor of
1.5 (this is also true for the masses of the first KK graviton and the second
KK gluon), which is larger the width of the KK-gluon.
Also, as we will discuss in section 4,  the angular
distribution of the decay products 
is different and should help distinguish the spin-1 and spin-2
modes. For these reasons, we have not included the effect of KK-gluon
in our study of discovery reach.

The channel for graviton decay to scalars (i.e. Higgs and longitudinal 
W's and Z's)  has a smaller branching ratio in the region 
 $\nu_{t,R} \gtrsim 1$, $\sim 30 \%$ in the model of Ref.~\cite{agashe}.
On the other hand, it could  be the  dominant mode 
in the region $-1/2 < \nu_{t,R} < 1$.  The existence of such a decay channel
could be very important in distinguishing KK-graviton channel from
KK-gluon production, since KK-gluon does not decay into such
states. Since the size of this channel is somewhat model dependent, we
will not take it into account in the following analysis. We expect it
will enhance the discovery potential for the KK-graviton in a generic
setup.

 Note that an important difference between a KK-gluon and
a heavy Higgs state, in addition to their spins, is the relative decay 
to weak gauge bosons. Both of them decay into $t \bar{t}$ as well as 
$WW$, $ZZ$. For the Higgs, due to the longitudinal enhancement and the
fact that the Higgs mass is proportional to its self-coupling, decays
into gauge bosons are dominant. However, there is no such enhancement
for the KK-graviton and $t \bar{t}$ tends  to be a somewhat larger
channel. 

\begin{figure}
\begin{center}
\includegraphics[angle=270,scale=0.35]{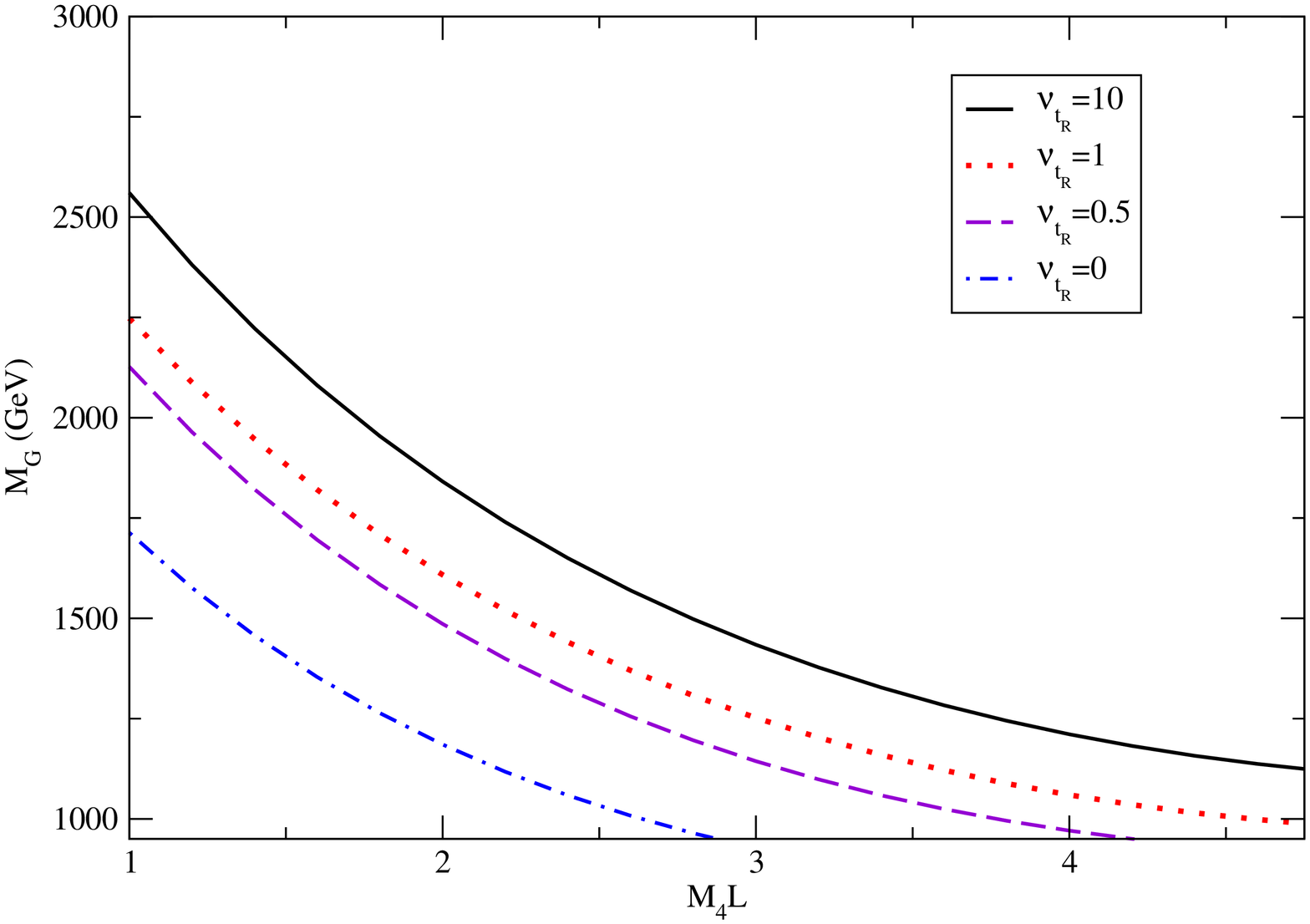}
\end{center}
\vspace{1cm}
\begin{center}
\includegraphics[angle=270,scale=0.35]{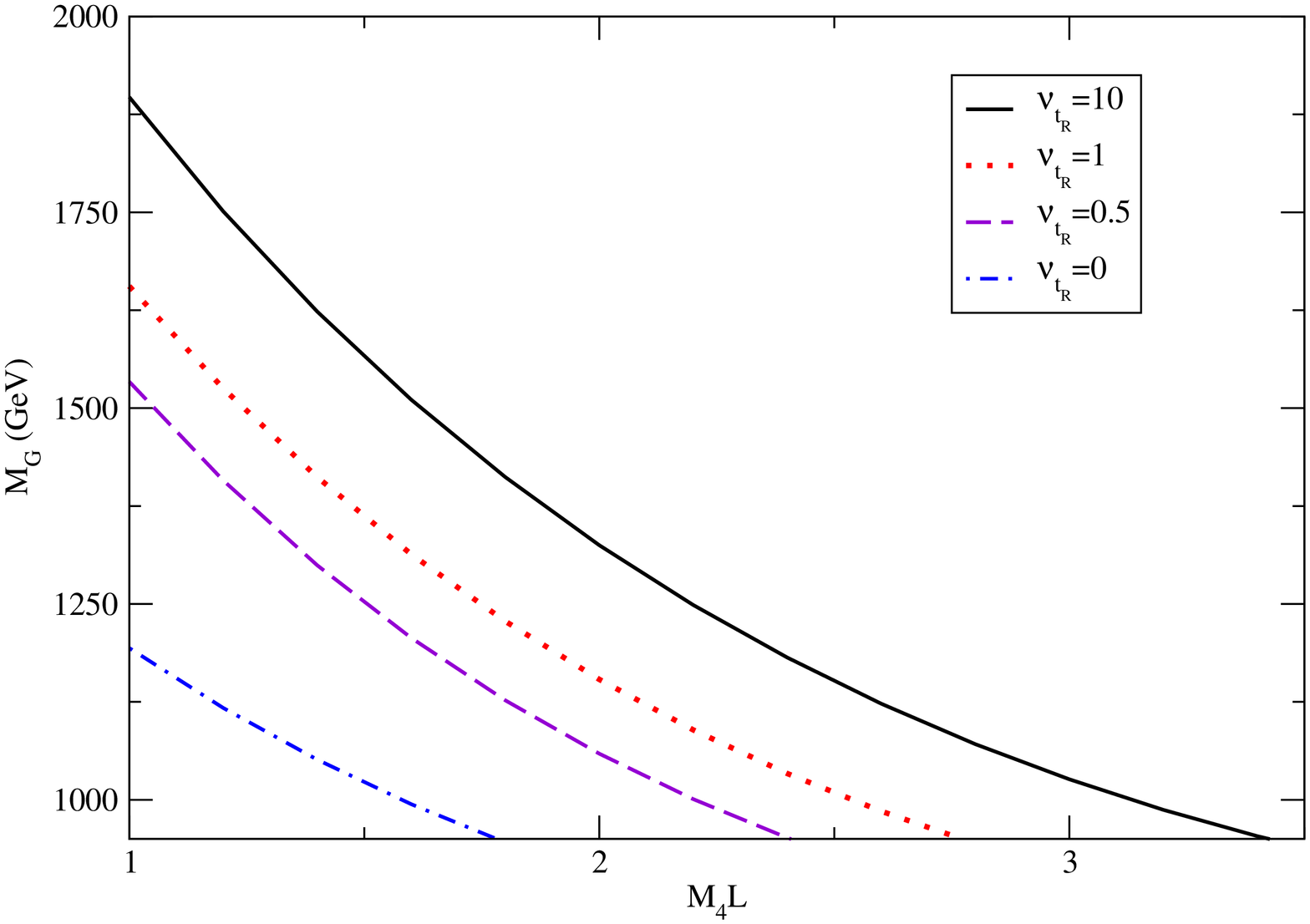}
\end{center}
\vspace{1cm}
\begin{center}
\includegraphics[angle=270,scale=0.35]{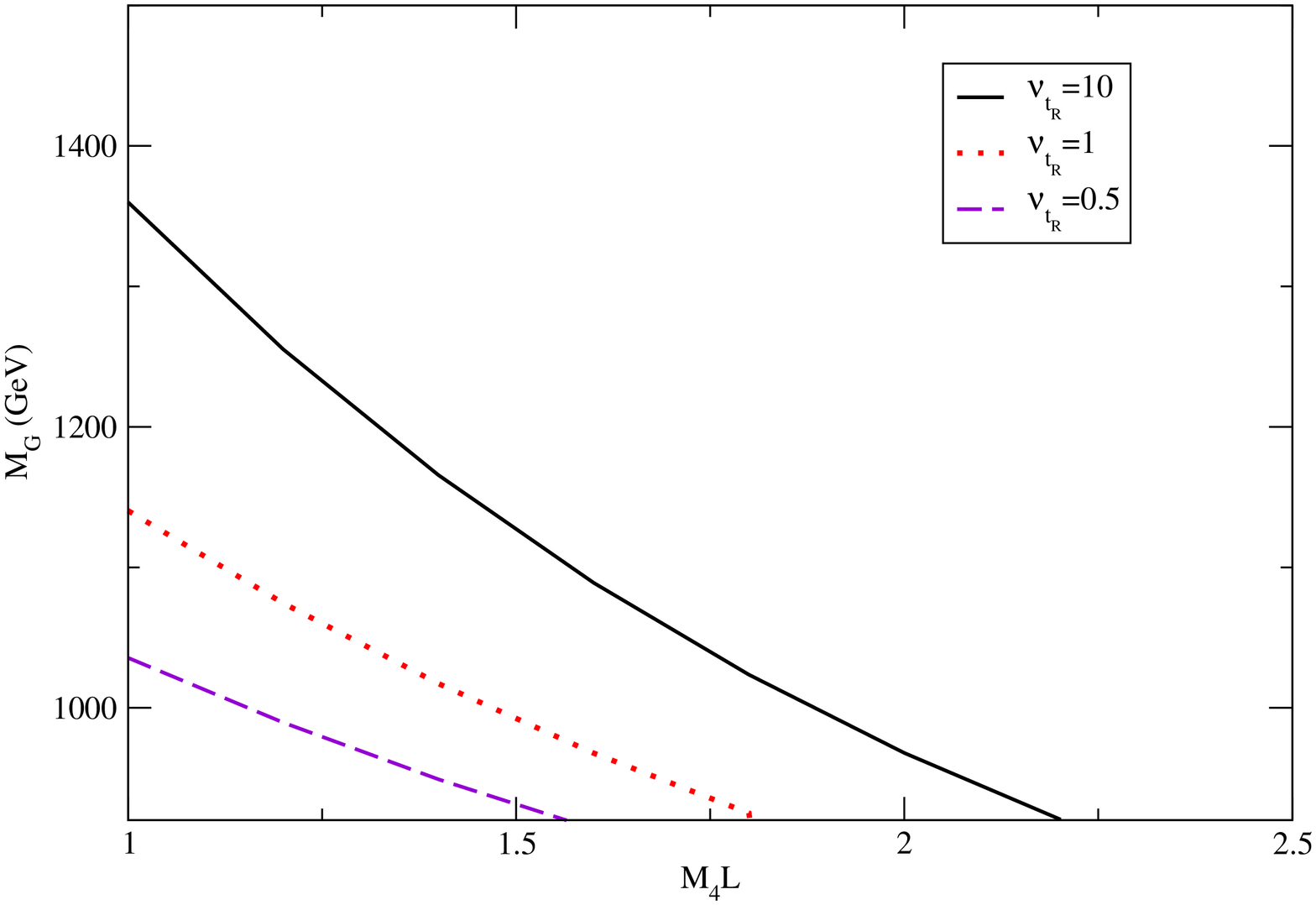}
\end{center}
\caption{The $s/\sqrt{b}=5$ reach as a function of graviton mass
and the parameter $(M_4 L)$. From top to bottom, the
reach is shown for 100\%, 10\%, and 1\%  efficient top identification.
Different levels of top IR localization are shown, $\nu_{t,R}=10.0,1.0,0.5,0.0$.
Larger $\nu_{t,R}$ corresponds to a more IR-localized $t_R$.  
\label{reach}}
\end{figure}

The rate for KK graviton production as function of the mass of the KK-graviton 
is shown in Fig.~\ref{xsec}.  Comparing with the SM $t \bar{t}$ background, we 
obtain the reach for discovery, shown in Fig.~\ref{reach}. For $\nu_{t,R}=0$,
the reach as a function of $M_4 L$ is roughly parameterized by $ m_{grav}=
2.8 TeV \times e^{-0.55 (M_4 L ) + 0.061 (M_4 L)^2}$, and
for $\nu_{t,R}=10.0$, by $ m_{grav}=
3.86 TeV \times e^{-0.45 (M_4 L ) + 0.040 (M_4 L)^2}$. There is no special
significance to the form of this parameterization.  We have assumed
100\% efficient top reconstruction.  The branching ratio to tops
decreases with decreasing $\nu_{t,R}$, so we have plotted the
reach for several possible values of $\nu_{t,R}$.  The KK-graviton
resonance is extremely narrow, even at small $M_4 L$, which cuts down on
the background.  However, the narrow resonance will be smeared out by
uncertainty in the measurement of the invariant mass of the KK resonance.
Thus, even with a very narrow width, the resonance will have to contend with
background events whose invariant mass is within a few percent of the graviton 
mass. We have estimated the effect of this uncertainty by taking 
the background to be all $t \bar{t}$ events within 3.0 \% $m_{grav}$
of the graviton mass, i.e., we have we used the smeared width as the
  window in which we compare signal vs background. The smearing we
have used $\sigma = E \times 3 \%$ is a typical value for ATLAS 
(\cite{atlasTDR}, Ch. 9).

\subsection{Energetic Top ID in $t \bar{t}$ final state}

Realistically, we will have to include top identification efficiencies
for both the signal and the background. Our signal significance,
assuming the efficiency for signal and background are roughly the
same, will scale down as the square root of the efficiency. Of course, any
top identification method will also introduce fakes from other
Standard Model processes, and these will effect the signal significance.  
A detailed study of these effects is beyond the scope of this paper.
In this section, we will discuss
new kinematical features of $t\bar{t}$ decaying from a heavy resonance, 
which will bring about uncertainties in top identification. We
comment on possible ways of developing alternative methods of
identifying tops in this type of processes. Due to such uncertainties
and potential room for improvement, we present the KK-graviton discovery
reach for a set of benchmark values of  top identifcation efficiencies.

Studies focused on identifying SM $t \bar{t}$ produced near threshold
typically yield a low efficiency \cite{atlasTDR}, \cite{cmsTDR},
\cite{Hubaut:2005er}, \cite{Baarmand:2006bn}, \cite{march}.
For example, in the ATLAS study, the combined efficiency 
of the semi-leptonic and pure leptonic channel is about $10 \%$
\cite{Hubaut:2005er}.
Since we are interested in a region which is far away from the $t \bar{t}$
production threshold, we expect the characteristics of the tops will
be different. Top identification in this kinematical regime is critical 
to KK graviton discovery.

\begin{figure}
\includegraphics{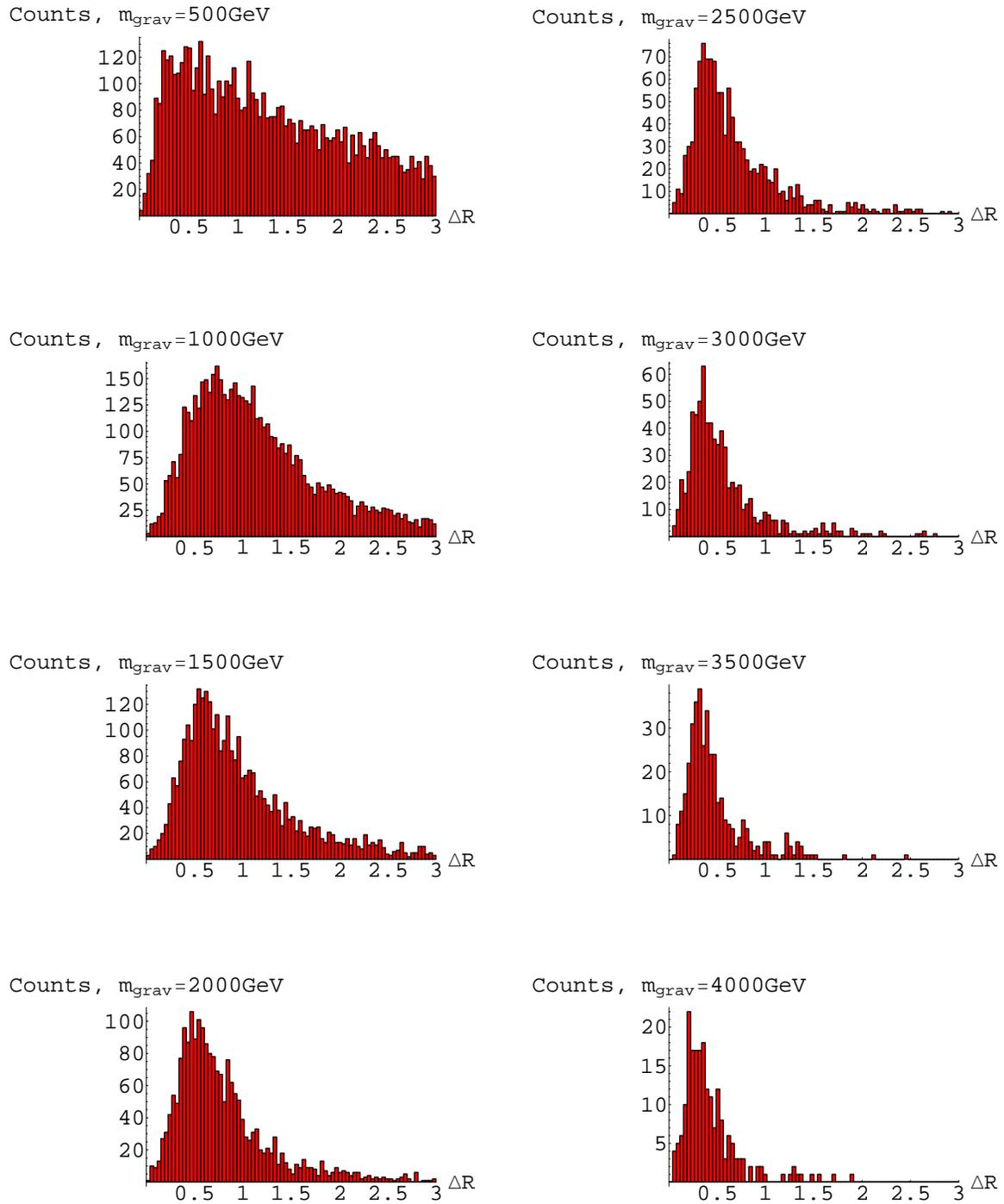}
\caption{Distributions of $\Delta R$ of top decay products for various KK masses.}
\label{deltar}
\end{figure}

The simplest method for top identification would be to construct the
invariant mass of the top quarks from their well separated decay products 
\cite{top_convention,atlasTDR,cmsTDR}. Doing this requires $\Delta R$, 
which measures the angle between the $b$ quark and lepton (for semileptonic
decay) or the maximal angle with the jet (for purely hadronic decay),
to be greater than 0.4 so that the final states are identified as
separate jets and a reasonably accurate invariant mass can be
calculated. In Figure \ref{deltar}, we show a Monte Carlo for the
expected $\Delta R$ for various values of the KK graviton mass. We see
that in all cases where we can hope to find the KK graviton resonance, 
a sizable fraction of the events have sufficiently large separation, 
which is very promising \cite{top-tao}.

This method might well suffice for top identification 
in the mass regime for which discovery is possible. To enhance 
statistics,  in addition to the conventional method, we can 
imagine other methods for top identification. 
In the case of a heavier KK-graviton,
we expect a sizable amount of the event will have one or two top quarks
highly collimated, especially if we use a somewhat larger cone size,
for example 0.7 for $m_{grav} \gtrsim 2.5 TeV$, see
Fig.~\ref{deltar}. If so, we expect they will show up in the form of 
one or two massive jets, which typically
have a lepton in them.  Without reliable
top identification to distinguish it from a QCD jet, we will have
to deal with a much bigger jet background that 
would make the collimated top quarks unobservable.  One possibility is
that we could use a massive jet algorithm, for example \cite{Wjmass},
so that all the decay products 
of each top that fall within the jet cone have a large invariant mass.  
However, QCD could also 
produce massive jets via off-shell partons and its contribution could
be significant \cite{Campbell:2006wx}. For this reason, an alternative method,
based on the different substructure of top jets and QCD jets could be useful.
Such substructure could be probed, for example, by
using finer granularity on the tracks which 
would provide additional information on the substructure of the
top-like objects. Given the importance of the energetic top signal, we
consider further detailed study on the experimental viability of such
a signature to be very worthwhile.  

There could be room for improvement in the top
identification efficiency. First of all, the top quarks produced from the
KK resonance will have significantly different kinematics
compared with the threshold production region considered in the studies
\cite{atlasTDR}, \cite{cmsTDR}, \cite{Hubaut:2005er}, \cite{Baarmand:2006bn}, 
\cite{march}. For example, the top quarks produced from
KK graviton decay will tend to be much harder ($p_T^t \sim 0.5
m_{\rm reson}$), and therefore less trigger suppressed.  On the other
hand, large boosts will tend to make the top decay products more
collimated. However, as can be seen from Figure 4, within the range of
mass we are interested in, this does not represents a significant
reduction in the signal rate.  

Notice that our focus is on identifying $t \bar{t}$ resonance,
rather than fully reconstructing all the decay products of top quark.
Therefore, one could also explore alternative strategies, such as applying a
looser definition of the top quark, including less focus on b-tagging and a 
relaxed W reconstruction condition. For highly boosted top quarks, one
could also attempt to improve the efficiency by identifying the tops as
massive fat jets with some substructure. Of course, all of these methods will
introduce new backgrounds which need to be included in the analysis.

Since the top identification efficiency is uncertain, in figure \ref{reach} we have 
plotted the KK-graviton discovery reach for efficiencies of $1 \%$ (the minimum 
expected), $10 \%$ (as in the quoted studies), and $100 \%$.
We see that for $1 \%$ efficiency, the reach is typically around
$m_{\textrm{grav}} \sim $ 1 TeV.
Results for any other efficiency could be obtained by scaling $s/\sqrt{b}$ by
$\sqrt{\frac{\textrm{eff}_1}{\textrm{eff}_2}}$.  Since the cross-section
approximately satisfies $\sigma \propto (M_4 L)^{-2}$, this can be compensated
for by a rescaling of $M_4 L \rightarrow M_4 L \left( \frac{\textrm{eff}_1}
{\textrm{eff}_2} \right)^{1/4}$.  For example, a reach of $m_{\textrm{grav}} = 2 $ TeV 
at $M_4 L = 1.8$ with 100\% efficient top identification corresponds to the same
reach at $M_4 L = 1$ with 10\% efficiency, as one can verify by inspection.

\section{Spin measurement}

\begin{figure}
\begin{center}
\includegraphics[angle=270,scale=0.5]{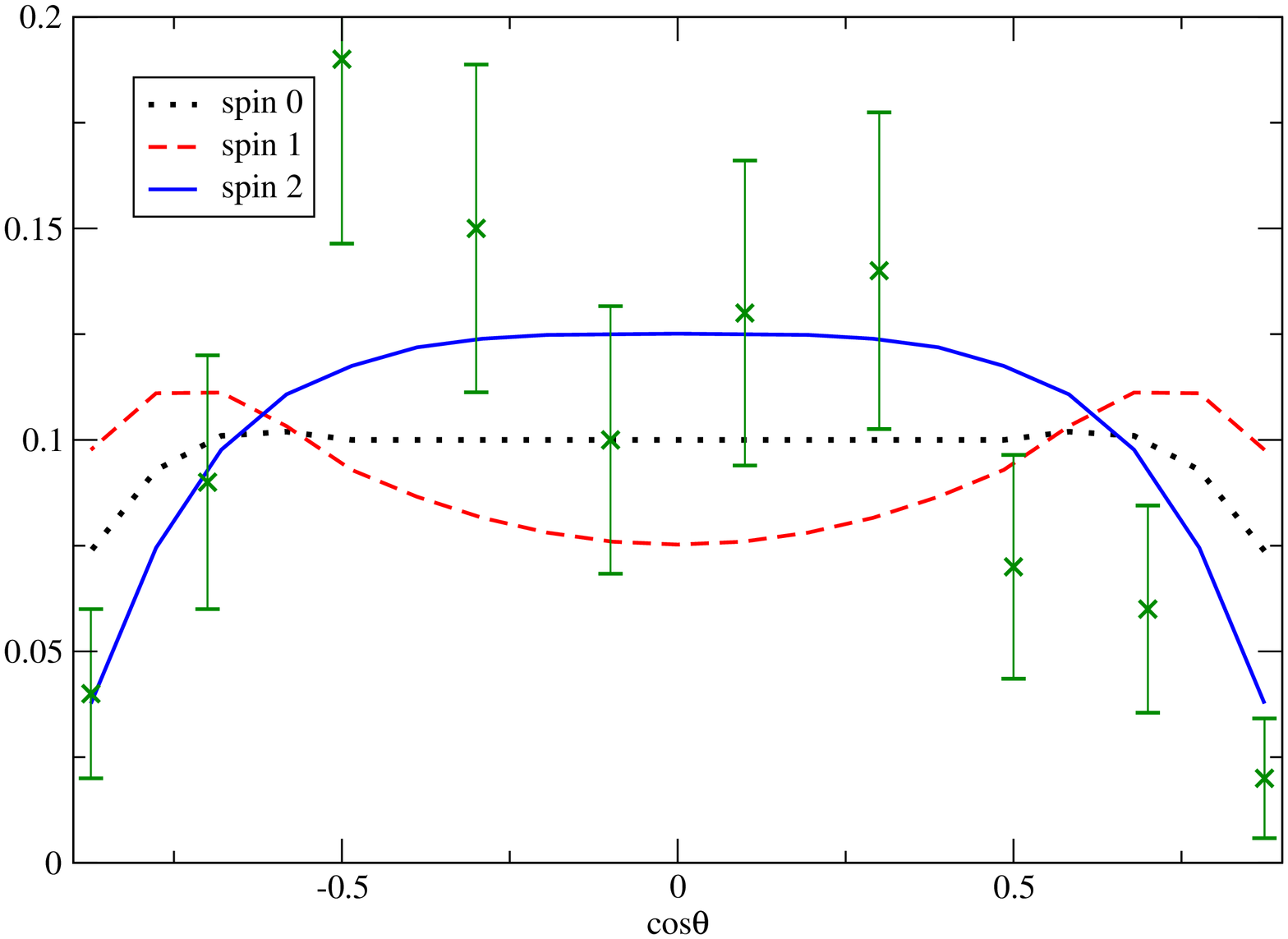}
\end{center}
\caption{Generic Sample of 100 $gg \rightarrow G \rightarrow t\bar{t}$ events.
The results of a Monte Carlo of 100 spin-2 events are shown on top of
the expected results for a spin-0, spin-1, and spin-2 resonance.  Events have 
been removed both from the expected curves and the Monte Carlo events
if they are too close to forward or backward scattering to be likely to be
observed (that is, if pseudo-rapidity $\eta > 2.5 - \ln 2$). }
\label{spin12evts}
\end{figure}

The channel $q \bar{q} \rightarrow V \rightarrow {t \bar{t}}$ will have
the characteristic distribution of $1+ \cos^2 \theta$ since it is
dominated by the transverse mode of vector boson $V$. On the other
hand, the KK-graviton, produced through gluon fusion, will  have a 
$1 - cos^4\theta $ dependence. This leads to a distinct difference from
a spin-1 resonance in the
cross-section near forward and backward scattering and could in principle
allow one to rule out a spin-1 particle with $\mathcal{O}(100)$ reconstructed 
top pairs.  A generic sample of 100 $t \bar{t}$ events, binned in 10 
bins from $\cos\theta = -1$ to $\cos\theta = +1$, is shown in Figure 
\ref{spin12evts}. The $\chi^2/ndof$ for the spin-2,spin-1, and spin-0 
distributions shown is 0.99, 3.7, and 2.1, respectively, where the 
number of degrees of freedom is $ndof=10$.
 The expected number of bins is shown for a spin-2, spin-1,
or spin-0 resonance.  The resolution is lower at near forward or
backward scattering, 
requiring a cut on pseudo-rapidity $\eta > 2.5$.  We have taken
this into account by conservatively assuming all gluons are as boosted as
kinematically allowed, so we cut out events with  $\eta > 2.5 - \ln 2$ in the
graviton rest frame.  As a result, some events have been cut from all three
distributions in the two extreme bins.  A spin-0 distribution is more difficult
to rule out than a spin-1 distribution, and would require more events.

The fact that the cross-section $\sigma_{gg \rightarrow G \rightarrow t\bar{t}}$
vanishes at $\cos \theta = \pm 1$ follows from conservation of angular momentum.
Of the five polarizations for the KK graviton,
only the three with $L_z = \pm 2, 0$ can be produced by the gluons.  
Moreover, the incoming and outgoing
state must have total angular momentum $l=2$, so the KK graviton
cannot decay to an s-wave top-pair, but instead must decay to a p-wave
top pair. The only p-wave spherical harmonic that does not vanish along
the z-axis is $Y_{10}$, so the spatial wave function cannot contribute
to the value of $l_z$ and thus the tops 
cannot couple to the $L_z = \pm 2$
polarizations.  Furthermore, the coupling $\bar{v}(p) \gamma^\mu u(p')$
vanishes when the quarks have the same helicity, so there is no
coupling to the $L_z=0$ polarization either.  Thus the 
total cross-section vanishes along the z-axis, as shown in Figure
\ref{spin12evts}.  

Note that the fact that the KK-graviton has spin 2 could be used to 
our advantage in top identification -- it may be possible to
cut down on background by using the angular distribution of the tops
to preferentially select the central region of phase space in figure
\ref{spin12evts}.

\section{Conclusion}

We have considered the LHC signatures of a KK graviton within the context
of RSI, when the top quarks are localized very near the TeV brane and the lighter
quarks are localized near the Planck brane.  We computed the 
cross-section for KK gravitons in this model and the discovery reach
from $t\bar{t}$ pairs.  We find that the KK graviton resonance is
very narrow, its width being less than a few percent of its mass 
even for $M_4 /k$ very close to 1,
which is distinctly different from the case when all fermions are localized
on the brane and there is a large number of possible decays for the graviton.
The dominant production/decay mechanism is gluons $\rightarrow $ graviton 
$\rightarrow t\bar{t}$, and in this case the angular distribution of the
cross-section is easier to distinguish from that of a spin-1 resonance
than in models with fermions on the brane.  The reason is that conservation of
angular momentum forces the cross-section to vanish at forward and backward
scattering for a spin-2 resonance, whereas the cross-section
increases at forward and backward scattering for a spin-1 resonance. We
find that the spin-2 distribution can be resolved with $\sim 100$ events.
In on-the-brane
models, the dominant production/decay mechanism is fermions $\rightarrow$
graviton $\rightarrow$ fermions, and the spin-2 nature of the graviton is
not as obvious.  In other models, where the left handed quark doublet of the
third generation is largely composite, both the rate of KK-graviton production
and its branching ratio to top and bottom quarks could be larger.  Thus 
KK-graviton phenomenology in these models deserves further study.

We find that the collider reach for detection is $m_{grav} \lesssim 1.7 TeV$, and
depends on the AdS curvature scale.  Detecting the graviton at the limits
of the collider reach depends on efficient reconstruction of the top decay
products, and further work is necessary to determine how efficiently this
can be done in practice.  Because the tops result from the decay of a 
very massive graviton, they will be highly boosted and thus their
decay products come out in a narrow cone (0.4 $\lesssim \delta R  \lesssim
2.0$, depending on the KK graviton mass). 

\section*{Acknowledgments}

We wish to thank B. Lillie, M. Franklin, Tao Han, Chris Tully for
useful discussions.  LR is supported by 
NSF grants PHY-0201124 and PHY-0556111.  ALF and JK are supported by
NSF graduate research fellowships, JK is also supported by a Hertz Fellowship,
and L.-T. W. is supported by DOE
under contract DE-FG02-91ER40654. This material is based upon work supported 
in part by the National Science Foundation under Grant No. 0243680. Any 
opinions, findings, and conclusions or recommendations expressed in this 
material are those of the author(s) and do not necessarily reflect the views 
of the National Science Foundation.

\section*{\textit{Note Added}}

Shortly after this work was submitted, there appeared the closely related
work \cite{agashegrav}, focusing on KK graviton decays to W's, 
and arguing that a slightly lower
limit on $M_4 L$ is allowed than that considered here.

\appendix

\section{Goldstone Boson Equivalence on the Brane}
\label{Goldstone}

The coupling between the W bosons and the graviton is difficult
to calculate in the 5d theory.  A precise calculation would
involve writing down the 5d equations of motion for the gauge
fields (which have no bulk mass term), writing down the symmetry-breaking
Higgs terms on the TeV brane, and solving for the wavefunctions and
mass eigenmodes satisfying the modified boundary conditions.  
Then, these would be integrated against the graviton wavefunction.
The rough picture for the 5d wavefunctions is that, before
symmetry-breaking, the gauge bosons start out with flat wavefunctions
and the Higgs starts out with a delta function wavefunction on the brane.
After symmetry-breaking, the gauge bosons eat a Higgs, and 
their wavefunctions are still mostly flat in the bulk but dip sharply
near the brane.

\begin{figure}
\begin{center}
\includegraphics{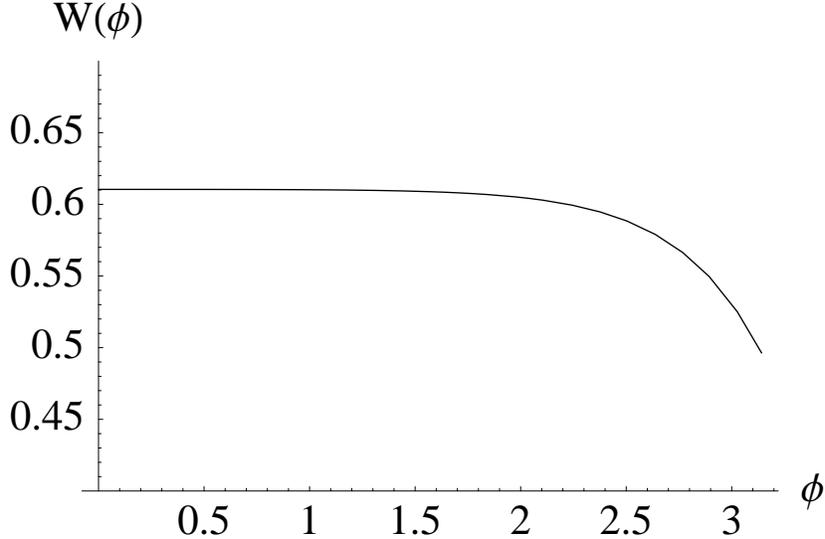}
\end{center}
\caption{When the Higgs gets a vev, the boundary conditions for the
wavefunction of the W boson are modified, and the wavefunction dips
down near the TeV brane.}
\end{figure}

A much easier and more transparent method is to go directly to the
effective KK theory before including the effects of symmetry-breaking.
Then, the W bosons pick up a mass from the usual Higgs mechanism
in 4d, and their couplings are determined from the Higgs couplings
by the usual Goldstone boson equivalence theorem, which states
\begin{equation}
\epsilon_{\mu_1} \epsilon_{\mu_2} \dots \epsilon_{\mu_n} 
\Gamma^{\mu_1\mu_2 \dots \mu_n}
   = \Gamma
\end{equation}
where $\Gamma^{\mu_1\mu_2 \dots \mu_n}$ is the vertex for $n$ 
W bosons, $\epsilon^{\mu_i}$ are the longitudinal polarization vectors,
and $\Gamma$ is the vertex for $n$ Higgses.  This can be seen
directly at the level of the lagrangian as well.  The Higgs
kinetic and interaction term are
\begin{equation}
\L \supset (\d H)^2 + c_{ssG} \d_\mu H \d_\nu H h^{\mu\nu}
\end{equation}
where $c_{ssG}$ is calculated using only the graviton wavefunction.
Gauge invariance then constrains the W interaction terms to arise by
promoting coordinate derivatives to covariant derivatives:
\begin{equation}
L \supset \left( (\d_\mu - i \frac{g_5\psi(\pi)}{2} W_\mu) H\right)^2 +
           c_{ssG} \left( \d_\mu H - i \frac{g_5\psi(\pi)}{2} W_\mu H \right)
                   \left( \d_\nu H - i \frac{g_5\psi(\pi)}{2} W_\nu H \right)
                    h^{\mu\nu}
\end{equation}
The factor $\psi(\pi)$ is the W wavefunction evaluated on the TeV brane.
We can absorb this factor into the coupling:
\begin{equation}
g_4 = g_5 \psi(\pi)
\end{equation}
If $v = \langle H \rangle$ is the Higgs vev, then the mass term for W
and the coupling to the graviton can be written
\begin{eqnarray}
\L &\supset& \left( \frac{g_4 v}{2} \right)^2 W^2 +
    c_{ssG} \left(\frac{g_4 v}{2} \right)^2 W_\mu W_\nu h^{\mu\nu} \\
c_{ssG} &=& \frac{2}{(M_4 L) \mu_{TeV}}
\end{eqnarray}
so it is manifest that the coupling of the W to the graviton is just $c_{ssG}$
times the mass-squared.  At high energies, the longitudinal polarization
$\epsilon^\mu  = \frac{p^\mu}{m_W} + \mathcal{O} \left( \frac{m_w}{E_p} \right)$,
so the W coupling to the graviton acts exactly as a Higgs.

\section{Loops}

In principle, there are contributions from top loops that could enhance
the production cross section for KK gravitons.  These contributions could
be relevant because the top coupling to the KK graviton is much stronger than
the gluon coupling, which is suppressed by a volume factor $\pi k r_c$.

The KK description of RS is a (Wilsonian) effective field theory
defined below the scale $N \mu_{TeV}$.
We can write the relevant interactions very schematically as
\begin{equation}
L \supset \frac{c_1}{N \mu_{TeV}} F_{\mu \rho} F_\nu^\rho h^{\mu \nu}
+ \frac{c_2}{N \nu_{TeV}}  h^{\mu \nu} i \psi^\dag  \bar{\sigma}_\mu D_\nu \psi
\end{equation}
To make predictions, we would in principle match this EFT to a UV
description at some matching
scale $\mu_{match}$, and then we would use the RG running of $c_1$ and
$c_2$ to make
predictions at any other scale (and we would also have other $c_i$'s
for higher order interactions).

However, in our case we do not have an accessible UV description.  All
we have is a tree-level
matching condition at an unknown matching scale, assumed to be of
order the cutoff.  This
seems to set a limit on the precision of our predictions.  For
instance, consider
the amplitude for two gluons and a KK graviton, which is schematically given as
\begin{equation}
A = \frac{p^2}{N \mu_{TeV}} \left( c_1(\mu)  + \frac{g^2 c_2(\mu)}{16
\pi^2} \right)
\end{equation}
where $\mu$ is the relevant scale.  Clearly if the $c_2$ contribution
is much smaller than the
contribution from $c_1$, then we should ignore it.  However, if
\begin{equation}
c_1(\mu_{match}) \ll  \frac{g^2 c_2}{16 \pi^2}
\end{equation}
then it would be very unnatural to ignore the loop correction from
$c_2$, because even if
$c_1$ is very small at some scale, it will be regenerated by $c_2$.
In our case this is especially
true, because we do not even know $\mu_{match}$.  Thus the loop
contribution sets a natural
lower bound on $c_1$.  Now if
\begin{equation}
c_1(\mu_{match}) \gtrsim  \frac{g^2 c_2}{16 \pi^2}
\end{equation}
then we can include the loop correction, but it does not seem to make
our analysis more precise.
This is because we only know $c_1$ at tree level, while there are
unknown corrections to it from
the one-loop matching at $\mu_{match}$ and from the RG running of the
coefficient $c_1$, and
these are of the same order as the loop.

In our case, we found that the top loop contribution is smaller than
the tree level
gluon contribution, although they are roughly of the same order of
magnitude.  This can
be viewed as an additional source of error in our analysis.

\section{Analytic Cross-section Estimates}

\subsection{W Boson Fusion}

We are interested in approximating the cross-section for 
$q_1 q_2 \rightarrow q'_1 q'_2 W W \rightarrow q'_1 q'_2 G$, where $G$ is 
a graviton\footnote{This section uses the techniques used in deriving
the effective W approximation, see \cite{cahnheavyhiggs}}.  
The fermion-fermion-W vertex is
$$
\frac{g_w}{\sqrt{2}} \bar{u}(q) \gamma^\mu \left( \frac{1-\gamma^5}{2} \right)
u(q') W_\mu
$$
and the  WWG vertex is
$$
\epsilon^{\mu\nu} W_\mu W_\nu m_w^2 c_{ssG}
$$
We work in the rest frame of the incoming quarks, so their 4-momentum is
$$ q_1 + q_2 = (2E, 0,0,0)$$
It is convenient to parameterize the outgoing quark energies and the difference in
their directions by
\beeqnar
E_1'& =& (1- \eta) E \nn\\
E_2' &=& (1-\zeta) E \nn\\
\cos \theta &=& - \hat{p}'_1 \cdot \hat{p}'_2
\en
Denoting the graviton momentum as $k^\mu$, its 3-momentum is
\beeqnar
\vec{l}^2 &=& (\vec{p}'_1 + \vec{p}'_2)^2 = E_g^2 - m_g^2
\en
and after some algebra, we find
\beeqnar
\cos\theta &=& 1 - \frac{2 (\eta \zeta  - m_g^2/4E^2)}{(1- \eta) (1-\zeta)}
\en
The kinematic bounds on $\eta $ and $\zeta$ are therefore
$\eta \zeta > m_g^2 /4E^2$ and $\eta + \zeta < 1 + m_g^2 / 4 E^2$.
We parameterize the  outgoing quark directions by
\beeqnar
\hat{q}'_1 &=& (\cos \alpha \cos\beta, \sin \alpha, - \sin \beta \cos \alpha)\nn\\
\hat{q}'_2 &=& -\hat{q}'_1 (\alpha \rightarrow \alpha - \theta)
\en
We denote the W momenta as $k_1 = q_1 -q'_1 $ and $k_2 = q_2 - q'_2$.  
The matrix element squared, averaged over initial spins and summed over final ones,
is therefore
\beeqnar
|M|^2 &=& \frac{g^4}{2} ( q_{1\mu} q'_{1\mu'} - q_1 \cdot q'_1 g_{\mu\mu'}
+ q_{1 \mu'} q'_{1\mu} ) ( q_{2\nu} q'_{2\nu'} - q_2 \cdot q'_2 g_{\nu\nu'}
+ q_{2 \nu'} q'_{2\nu} ) \nn\\
  &\times & \left( \frac{1}{k_1^2 - m_w^2 } \right)^2
  \left( \frac{1}{k_2^2 - m_w^2 } \right)^2  D_{\mu\nu; \mu'\nu'}
\en
where $D_{\mu\nu; \mu'\nu'}$ is the sum over graviton polarizations, equal
to the numerator of the graviton propagator:
\beeqnar
D_{\mu\nu; \mu'\nu'} &=& \half \left( G_{\mu\mu'} G_{\nu\nu'} 
 + G_{\mu\nu'} G_{\mu'\nu} - \frac{2}{3} G_{\mu\mu'} G_{\nu\nu'} \right) \nn\\
G_{\mu\nu} &=& g_{\mu\nu} - \frac{k_\mu k_\nu}{m_g^2}
\en
The W propagators can be simplified:
\beeqnar
k_1^2 - m_w^2 &=& 2 E^2 (1-\eta) \left( x + \cos \alpha \sin\beta\right)
\nn\\
k_2^2 - m_w^2 &=& 2 E^2 (1-\zeta) \left( y + \cos(\alpha - \theta)
\sin\beta \right) \nn\\
x &=& 1 + \frac{m_w^2}{2 E^2 (1-\eta) }\nn\\
y &=& 1 + \frac{m_w^2}{2 E^2 (1-\zeta)} 
\en
When $E \gg m_w$, there is a near-divergence in the W propagators that
is cut off by the W mass, and the dominant contribution to the cross-section
comes from $\theta \sim 0$ and $\cos\alpha \sin \beta = -1$. 
 We can expand $\alpha$ around $\pi$ and $\beta$ around $\pi/2$:
\beeqnar
\alpha &=& \pi - \nu \nn\\
\beta &=& \pi/2 - \epsilon \nn\\
\cos \alpha &\sim& -1 + \half \nu^2 \nn\\
\sin \beta &\sim& 1 - \half \epsilon^2 \nn\\
\cos ( \alpha - \theta) &\sim& -1 + \half (\nu + \theta)^2
\en
We are left with an expression that is still relatively complicated. We 
start by focusing on the $\nu-$,$\ep-$, and $\theta$-dependence:
\beeqnar
d \sigma &\propto& J(\theta) =
16 \int d\nu d\epsilon \left( \frac{ 1 + \O(\nu^2,\epsilon^2) }
{2(x-1) + (\nu^2 + \epsilon^2)}\right)^2 
\left( \frac{1 + \O(\nu^2,\ep^2)}{2(y-1) + ((\nu+\theta)^2  + \ep^2)} \right)^2
\en
The leading order term can be evaluated exactly.  The $\O(\nu^2,\ep^2)$
terms diverge individually (of course, they are part of an
expansion of $\cos$ so they do not diverge if summed up) so we will drop them.
They would lead to a partial cancellation, since we are essentially
approximating $\cos\nu= 1 - \half \nu^2 + \dots$ by 
$\cos \nu = 1$ near $\nu = 0$ in the numerator. This amounts to keeping
the angular dependence of only the $(k_1^2 - m_w^2)^{-2} (k_2^2 - m_w^2)^{-2}$
piece in the amplitude.  These are the terms responsible for the 
divergence (in the limit $E \gg m_w$) when the W's are collinear.
Consequently,
the remaining terms get their dominant contribution from $\theta=0$,
so we approximate  $J\propto \delta(\theta^2)$, with proportionality contant
given by
\beeqnar
\int d \theta^2 J 
&=& 2 \int \theta d\theta J = \frac{1}{\pi} \int d^2 \theta J \nn\\
  &=& \frac{16}{\pi} \int 
\frac{d\nu d\ep}{(2 (x-1) + (\nu^2 + \epsilon^2))^2} 
\int \frac{d \theta_x d\theta_y }
 {(2(y-1) + ((\nu+\theta_x)^2  + (\ep+\theta_y)^2)^2} \nn\\
  &=& \frac{4\pi}{(x-1) (y-1)}
\en
So, in this approximation,
\beeqnar
J &=& \int d\alpha d\cos\beta \left( \frac{1}{k_1^2 - m_w^2} \right)^2 
\left( \frac{1}{k_2^2 - m_w^2} \right)^2 \nn\\
  &=&\frac{4 \pi}{(x-1)(y-1)} \delta(\theta^2)
\en
The rest of $d \sigma$ can now be evaluated at this level of approximation
by a straightforward but tedious computation,
since $\alpha = \pi, \beta = \pi/2$ and $\theta = 0$:
\beeqnar
(k_1^2 - m_w^2)^2(k_2^2 -m_w^2) d \sigma &=& 
  \frac{16}{3} E^4 (1- \zeta)(1-\eta) (1 - \zeta \eta + \zeta^2 \eta^2) 
\times \left( g^4 2^{-10} \pi^{-4} m_w^4 c_{ssG}^2 \right)
\en
Putting everything together and using $\theta^2 =
\frac{4 (\eta \zeta - m_g^2/4E^2)}{ (1-\eta)(1-\zeta)} $ from above, we get
\beeqnar
\sigma &=&
\int_{m_g^2/4E^2}^1 d\eta \int_{m_g^2 /4E^2}^{1+(m_g^2/4E^2) - \eta} d\zeta
\frac{g^4}{2^{12} \pi^4} \frac{16 \pi}{3} \frac{1}{m_w^4} 
  (1 - \zeta \eta + \zeta^2 \eta^2 ) 
\delta (\frac{4 (\eta \zeta - m_g^2/4E^2)}{ (1-\eta)(1-\zeta)} m_w^4 c_{ssG}^2
\nn\\
&=& \frac{g^4}{(3) 2^{10} \pi^3} c_{ssG}^2 (1 - \hat{\tau} + \hat{\tau}^2)
\left( (1 + \hat{\tau}) \log \left( \frac{1}{\hat{\tau}} \right) 
- 2 ( 1 - \hat{\tau}) \right)
\en
where $\hat{\tau} = \frac{m_g^2}{4 E^2} = \frac{m_g^2}{\hat{s}}$. This
contains the usual luminosity function 
$-((1+\hat{\tau})\log(\hat{\tau}) + 2 (1- \hat{\tau}))$ 
for effective longitudinal W's, as well as an additional piece from
the spin structure of the resonance.
We convolve this with the fermion luminosity function
\beeqnar
\dd{\L}{\tau} &=& 2 \int_\tau^1 \frac{dx}{x} f_+(x) f_-(\tau/x) d\tau
\en
where $f_+$ is the sum of pdfs for the positively charged fermions and $f_-$
is for the negatively charged ones. The cross-section is then
$$
\int_{\hat{\tau}}^1 \dd{\L}{\tau} (\tau) \sigma (\tau)
$$
To evaluate this numerically, we used the CTEQ5M parton distribution
functions in their Mathematica distribution package\cite{cteq}. 
We compare this with the results of the Monte Carlo integration below:

\begin{tabular}{|lll|lll|}
\hline
$m_{grav}$ (TeV) & $\sigma_{est} $ (fb) & $\sigma_{prog}$ (fb) &
$m_{grav}$ (TeV) & $\sigma_{est} $ (fb) & $\sigma_{prog}$ (fb) \\
\hline
0.5 &  47  & 62 & 2.5 & 0.012 &  0.0089 \\
1.0 &  2.3 &  2.3 & 3.0 & 0.0029 &  0.0022 \\
1.5 & 0.29 &  0.25 & 3.5 & 0.00079 &  0.00056 \\
2.0 & 0.053 &  0.042 & 4.0 & 0.00022 & 0.00016 \\
\hline
\end{tabular}

For the convenience of the reader, we found that the CTEQ5M pdf's at 
an energy scale $Q=200 TeV$ can be parameterized by
\begin{eqnarray*}
x g(x) &=& \frac{ 1.549 e^{-3.113 x^{1/3}} (1-x)^{5.448}}{x^{0.5270}} \\
x u(x) &=& \frac{0.02596 e^{7.667 \sqrt{x}} (1-x)^{7.273}}{x^{0.6125}} \\
x d(x) &=& \frac{0.04735 e^{4.309 \sqrt{x}} (1-x)^{6.203}}{x^{0.5267}} 
\end{eqnarray*}

\subsection{Gluon Fusion}

Since the gluon fusion is a two-to-two decay, it can be evaluated exactly.
The amplitude is 
\beeqnar
\M &=& T_{gluon}^{\mu\nu} D_{\mu\nu;\mu'\nu'} T^{\mu'\nu'}_{top}
\en
This implies that the cross-section is
\beeqnar
\frac{d\sigma}{d \cos \theta} &=& 
   \frac{5}{64} \frac{m_g^4}{\hat{s}} ( 1- \cos^4 \theta) c_{AAG}^2 
  \frac{m_g^{-3} \Gamma_{t\bar{t}} }{ \left( \frac{\hat{s}}{m_g^2} - 1 \right)^2 
+ \frac{\Gamma_{\rm tot}^2}{m_g^2} }
\en
Integrating over $\cos\theta$, we find
\beeqnar
\sigma &=& \frac{1}{8} \frac{ \tau}{\hat{\tau}} 
  \left( \frac{0.47}{(35)(2.5)(m_g/3.83)} \right)^2
\frac{m_g \Gamma_{t\bar{t}}}{ \left( \frac{\hat{\tau}}{\tau} - 1\right)^2
  + \frac{\Gamma^2}{m_g^2} }
\en
To get the full cross-section, we convolve this with the
gluon luminosity function:
\beeqnar
\sigma_{tot} &=& \int_{\hat{\tau}}^1 
  \dd{\L}{\tau} \sigma(\hat{\tau},\tau) d\tau
\en
Comparison with the Monte Carlo integration is shown below
(setting $\Gamma_{t\bar{t}} = \Gamma_{\rm tot}$, so this calculates
the total graviton production cross-section, without the branching
ratio of subsequent decay to tops):

\begin{tabular}{|lll|lll|}
\hline
$m_{grav}$ (TeV) & $\sigma_{est} $ (fb) & $\sigma_{prog}$ (fb) &
$m_{grav}$ (TeV) & $\sigma_{est} $ (fb) & $\sigma_{prog}$ (fb) \\
\hline
0.5 &  2500  & 1300 & 2.5 & 0.32 &  0.40 \\
1.0 &  85 &  91 & 3.0 & 0.078 &  0.098 \\
1.5 & 9.0 &  11. & 3.5 & 0.021 &  0.026 \\
2.0 & 1.5 &  1.8 & 4.0 & 0.0061 & 0.0077 \\
\hline
\end{tabular}

\end{document}